\documentclass[dvips]{article}

\usepackage{icrc2011}
\usepackage{amssymb}

\title{MAGIC and Multi-Wavelength Observations of Mrk\,180 and 1ES\,2344+514 in 2008}
\newcommand{\etal}{\MakeLowercase{\textit{et al.}}} % "et al."
\shorttitle{R\"ugamer \etal: MW Observations of Mrk\,180 and 1ES\,2344+514}

\authors{R\"ugamer, S.$^{1}$, Angelakis, E.$^{2}$, Bastieri, D.$^{3}$, Dorner, D.$^{4}$, Fuhrmann, L.$^{2}$, Kovalev, Yu.\ A.$^{5}$, Kovalev, Y.\ Y.$^{2,5}$, L\"ahteenm\"aki, A.$^{6}$, Lindfors, E.$^{7}$, Longo, F.$^{8}$, Lucarelli, F.$^{9}$, Pittori, C.$^{9}$, Reinthal, R.$^{7}$, Sbarra, C.$^{3}$, Sokolovsky, K.\ V.$^{2,5}$, Stamerra, A.$^{10}$, Ungerechts, H.$^{11}$ for the MAGIC collaboration, the F-GAMMA program and the $Fermi$-LAT, RATAN, MOJAVE, AGILE and IRAM teams}
\afiliations{
$^1$Universit\"at W\"urzburg, D-97074 W\"urzburg, Germany\\
$^2$Max-Planck-Institut f\"ur Radioastronomie, D-53121 Bonn, Germany\\
$^3$Universit\`a di Padova and INFN, I-35131 Padova, Italy\\
$^4$ISDC Data Center for Astrophysics, CH-1290 Versoix, Switzerland\\
$^5$Astro Space Center of Lebedev Physical Institute, 117997 Moscow, Russia\\
$^6$Aalto University Mets\"ahovi Radio Observatory, FI-02540 Kylm\"al\"a, Finland\\
$^7$Tuorla Observatory, University of Turku, FI-21500 Piikki\"o, Finland\\
$^8$Universit\`a di Trieste, and INFN Trieste, I-34127 Trieste, Italy\\
$^9$Agenzia Spaziale Italiana (ASI) Science Data Center, I-00044 Frascati, Italy\\
$^{10}$Universit\`a  di Siena, and INFN Pisa, I-53100 Siena, Italy\\
$^{11}$Instituto de Radio Astronoma Milim\'{e}trica, E-18012 Granada, Spain
}
\email{snruegam@astro.uni-wuerzburg.de}

\abstract{
Simultaneous multi-wavelength (MW) campaigns are the most promising approaches to investigate the still unrevealed nature of blazars, active galactic nuclei which are variable on all time scales from radio to TeV energies.\\
In 2008, two MW campaigns on the high-frequency peaked blazars Mrk\,180 and 1ES\,2344+514 have been organised by the MAGIC collaboration. From radio to TeV gamma rays, RATAN-600, Mets\"ahovi, Effelsberg, VLBA (only 1ES\,2344+514), IRAM, KVA, $Swift$, AGILE, $Fermi$-LAT and MAGIC-I were taking part in these campaigns. Mrk\,180 had just been discovered at TeV energies by MAGIC in 2006, whereas 1ES\,2344+514 is a known TeV emitter since many years. Due to their rather faint emission particularly at TeV energies, the campaigns represented quite challenging observations. In fact, Mrk\,180 has not been investigated until now in MW campaigns, and for 1ES\,2344+514 only one campaign including TeV measurements has been reported in literature up to now.\\
In this contribution, we will present detailed MW light curves for both sources and describe the composite wide range spectral energy distributions by theoretical models.
}
\keywords{BL Lacertae objects - Mrk\,180 - 1ES\,2344+514 - multi-wavelength campaigns - VHE observations}

\begin{document}
\maketitle

%Begin the section.
\section{Introduction}
%The gamma-ray sky is dominated by blazars, radio-loud active galactic nuclei (AGN) whose non-thermal relativistic plasma jet is observed at small angles with respect to our line of sight. Their spectral energy distribution (SED) spans from the radio band up to the TeV energy regime and is characterised by two broad peaks. The first peak is produced by synchrotron radiation, whereas the origin of the second peak is still a matter of debate. The most common explanation involves inverse-Compton upscattering of the synchrotron photons (synchrotron self-compton models, SSC; e.g.\ \cite{SSC1, SSC2}) or photons from an external radiation field (external Compton, e.g.\ \cite{EC1, EC2}) to the gamma-ray regime. The SED can also be described by hadronic models (e.g.\ \cite{PIC1, PIC2}), though.

%BL Lacertae objects constitute a special class of blazars, showing no or only faint emission lines and their emission being dominated by non-thermal radiation. They are further subdivided according to the frequency of the synchrotron peak \cite{Padovani}: for high-frequency peaked BL Lacertae (HBL) objects, the peak is located in the UV to soft X-ray regime, and the second emission component peaks in the GeV to TeV range, which makes these objects prime targets for very high energy (VHE, $\gtrsim$ 100 GeV) observations.

Blazars are a subclass of active galactic nuclei dominated by non-thermal emission spanning from the radio band up to the TeV energy regime. Their spectral energy distribution (SED) is characterised by two broad peaks whose origin is most commonly ascribed to synchrotron and self-Compton (SSC) emission (e.g.\ \cite{SSC1, SSC2}; but see also different approaches, e.g.\  external Compton \cite{EC1, EC2} or hadronic models \cite{PIC1, PIC2}). They are further subdivided according to the frequency of the first peak \cite{Padovani}: for high-frequency peaked BL Lacertae (HBL) objects, the peak is located in the UV to soft X-ray regime, and the second emission component peaks in the GeV to TeV range, which makes these objects prime targets for very high energy (VHE, $\gtrsim$ 100 GeV) observations.

Blazars show variability at all wavelengths in flux as well as spectral shape on time scales from month down to minutes (e.g.\ \cite{HESS_2155, MAGIC_Mrk501}). Consequently, the SED %, one of the most important handles in assessing the physics of the source,
can only be measured meaningfully by means of simultaneous observations at all involved energy bands. Due to the organisational effort and low sensitivity especially of the former gamma-ray observatories, those campaigns were available only for a handful of sources at the time of 2008, and centered on the brightest objects or high flux states. The campaigns reported here (see also \cite{RihoICRC}) were organised intentionally regardless of the sources' flux state, and the chosen targets were hardly studied in MW campaigns.

Mrk\,180 is a nearby (z = 0.046) HBL discovered at VHE by MAGIC in 2006 following an optical high state \cite{Mrk180_MAGIC}. %The number of VHE detections during optical flares is increasing (e.g.\ \cite{MAGIC_1011}, \cite{MAGIC_0716}, \cite{MAGIC_1215}), suggesting a correlation between these two energy bands.
No second VHE detection has been published up to now. EGRET did not report a detection of the source \cite{EGRET_catalog}, but it is included in the $Fermi$-LAT 1-year Catalog \cite{Fermi_1year}. Flux variations in the optical are modest due to the strong host galaxy, outshining the core \cite{Mrk180_optical}. In X-rays, variability by $\sim$ a factor 3 at 1 keV have been reported \cite{Mrk180_Xrays}. The campaign described here represents the first multi-wavelength (MW) campaign including VHE observations for this object.

1ES\,2344+514, an HBL located at a redshift of z = 0.044, has been discovered in the VHE regime by Whipple in 1995 \cite{2344_Whipple}. Though being the third extragalactic VHE source to be detected, due to its faintness only one MW campaign including a VHE detection has been published up to now \cite{2344_MWL}. The source is listed in the $Fermi$-LAT 1-year Catalog \cite{Fermi_1year}, but EGRET did not detect the source \cite{EGRET_catalog}. In 2000, BeppoSAX reported strong flux changes on time scales of $\sim$ 5000 s and spectral changes in X-rays. During the highest flux state, the synchrotron peak was detected at energies $>$ 1 keV for this object \cite{2344_extreme}, adding it to the small but recently growing list of so-called 'extreme blazars' \cite{extreme_blazars}. Also this source shows only small flux variations in the optical due to a bright host galaxy \cite{2344_optical}.

\section{Observation Campaigns and Results}
\subsection{Mrk\,180}
The MW observations of Mrk\,180 were centered on common observation windows of the AGILE satellite and the MAGIC-I telescope \cite{MAGIC-I}. Two suitable observation windows were found, one in spring, the other one at the end of the year. Observations in the optical were conducted by the KVA telescope, which is operated concurrently with the MAGIC telescopes. Additionally, RATAN-600, Mets\"ahovi, Effelsberg and IRAM were performing snapshots of the source. The X-ray coverage was provided by ToO proposals to $Swift$. Finally, the second observation window was complemented by $Fermi$-LAT observations.% For details see Table \ref{tab:obs}.
%\begin{table*}
%\begin{center}
%\begin{small}
%\begin{tabular}{l||c|c|c}
%& Band & Mrk\,180 & 1ES\,2344+514\\\hline\hline
%RATAN-600 & 4.8 - 22.2 GHz & 10/01 - 02 & 09/20 - 10/03\\\hline
%Mets\"ahovi & 37 GHz & 03/22, 04/23 - 24, 05/21 - 23 & ---\\\hline
%Effelsberg & 2.6 - 43.0 GHz & 02/18, 11/08 & 10/17, 11/08, 12/06, 09/01/24\\\hline
%VLBA & 4.6 - 43.2 GHz & --- & 10/23\\\hline
%IRAM &  86.2 - 228.4 GHz & 02/11, 03/10, 05/05 \& 25, 06/26, 07/27, 08/25, 10/08 & ---\\\hline
%KVA & R, B, V & 02/23 - 06/03; 11/18 - 12/12 & 10/14 - 09/01/03\\\hline
%$Swift$ & UV \& X-ray & 04/29 - 05/10; 10/22 - 12/10 & 10/06 - 11/14\\\hline
%AGILE & HE & 04/30 - 05/10; 10/31 - 11/30 & 10/31 - 11/30\\\hline
%$Fermi$ & HE & 08/04 + & 08/04 +\\\hline
%MAGIC-I & VHE & 04/28 - 05/10; 10/23 - 12/08 & 10/20 - 11/30\\\hline
%\end{tabular}
%\caption{Multi-wavelength observations of Mrk\,180 and 1ES\,2344+514, conducted in 2008 unless stated otherwise. In the case of KVA, $Swift$ XRT and MAGIC-I, the given observation periods were not covered continuously.}\label{tab:obs}
%\end{small}
%\end{center}
%\end{table*}

\begin{figure}[!t]
\centering
\includegraphics[clip, width=3.2in]{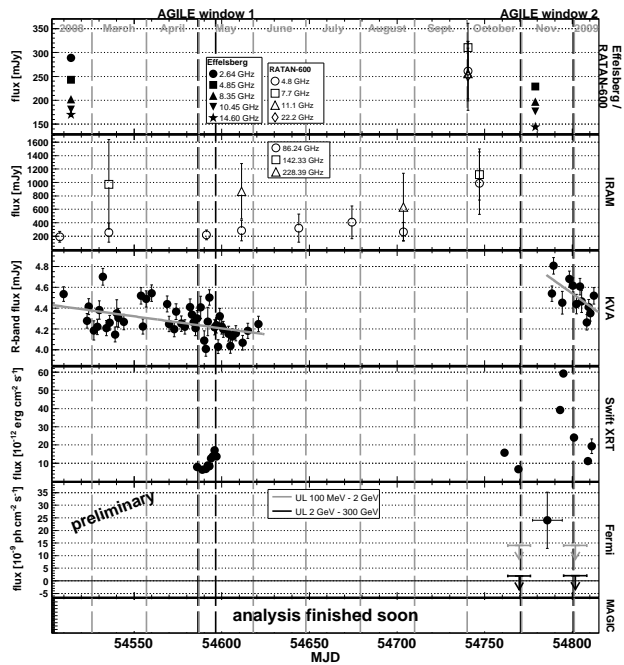}
\caption{MW light curve for Mrk\,180 with daily binning. The optical data points are not host-galaxy corrected. The grey solid lines represent linear fits to the optical data.}
\label{fig:LC_Mrk180}
\end{figure}

The light curves in Figure \ref{fig:LC_Mrk180} show seemingly uncorrelated trends. At radio wavelengths, the Effelsberg data pointed to a constant or decreasing flux whereas the RATAN-600 measurements indicated a higher flux state in between the Effelsberg measurements, though not significantly due to the large error bars. Also the IRAM data were consistent with a constant flux although showing a trend of higher flux at the time of the RATAN-600 observations, which was unfortunately not covered at other wavelengths. The optical flux was governed by irregular flux variations due to systematic uncertainties in the determination of the thermal background, but was also showing an overall decreasing trend within the first observation window. Also in the second window, the flux was decreasing albeit an overall higher flux compared to the first window was observed. In X-rays, XRT could clearly detect flux variability; within the first window, the flux was almost constantly rising, in total by $\sim$ a factor of 2. In the second window, the flux was going down first to increase in the course of a giant flare by a factor $\sim$ 9 and eventually to decrease again. During the second flare, the $Fermi$-LAT 1-year Catalog public light curve\footnote{http://fermi.gsfc.nasa.gov/ssc/data/access/lat/1yr\_catalog} also showed an increase in flux. Although the optical light curve seemed not to follow the short-term trends in X-rays, a short flare $\sim$ 3 days before the first X-ray flare could be observed. Due to the low sampling a correlation could not be established, though. The source could not be detected within the given time windows by Mets\"ahovi and AGILE.

\begin{figure}[!t]
\centering
\includegraphics[clip, width=3.2in]{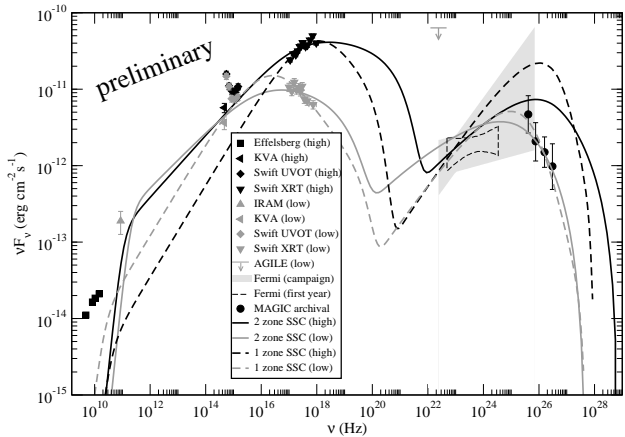}
\caption{SEDs of Mrk\,180 resulting from this campaign. The KVA data are host-galaxy corrected, the UVOT data are de-reddened but the thermal contribution is not subtracted. At high gamma-ray energies, a 95 \% c.l.\ upper limit derived by AGILE is shown together with $Fermi$ bow-ties deduced from the LAT 1-year Catalog (dashed line) and from the MAGIC-simultaneous period MJD 54763 - 54808 (grey shaded).}
\label{fig:SED_Mrk180}
\end{figure}

For MJDs 54590.9 and 54794.9, two $Swift$ nights with good MW coverage and large flux difference, simultaneous SEDs were derived as shown in Figure \ref{fig:SED_Mrk180}. With the MAGIC analysis not being finished yet, VHE points from the MAGIC discovery \cite{Mrk180_MAGIC} are shown for comparison% (corrected for extragalactic background light (EBL) absorption effects using the best-fit model of \cite{Kneiske_bestfit}). In the HE regime, a 95 \% c.l.\ upper limit derived by AGILE is shown together with $Fermi$ bow-ties deduced from the LAT 1-year Catalog (dashed line) and from the MAGIC-simultaneous period 10/24 - 12/08 (grey shaded). The $Swift$ UVOT data have been dereddened but not corrected for the thermal component of the source. The host-galaxy has been subtracted in the case of the KVA points.
.

\subsection{1ES\,2344+514}
For 1ES\,2344+514, %which is included in the AGILE team source list, only one suitable common observation window of AGILE and MAGIC was found. T
the same wavelength coverage as for Mrk\,180 was achieved and complemented by VLBA observations. %through the MOJAVE monitoring project. %The observation times are listed in Table \ref{tab:obs}. 
Though having detected the source about three month prior to this campaign, Mets\"ahovi did not detect 1ES\,2344+514 within the given time windows, excluding major flares to have happened at 37 GHz during the campaign. AGILE and $Fermi$ did not detect the source within the analysed observation windows (MJDs 54770 - 54800 and 54730 - 54830, respectively).

The analysis of the MAGIC-I data yielded a significance of $>$ 3 $\sigma$, which is below the standard VHE detection significance of 5 $\sigma$. Taking into account the long observation time of $\sim$ 20 h and 1ES\,2344+514 being a well-established VHE source, we derived an average spectrum of the complete data set. Nevertheless the low significance should be kept in mind when interpreting the results.

\begin{figure}[!t]
\centering
\includegraphics[clip, width=3.2in]{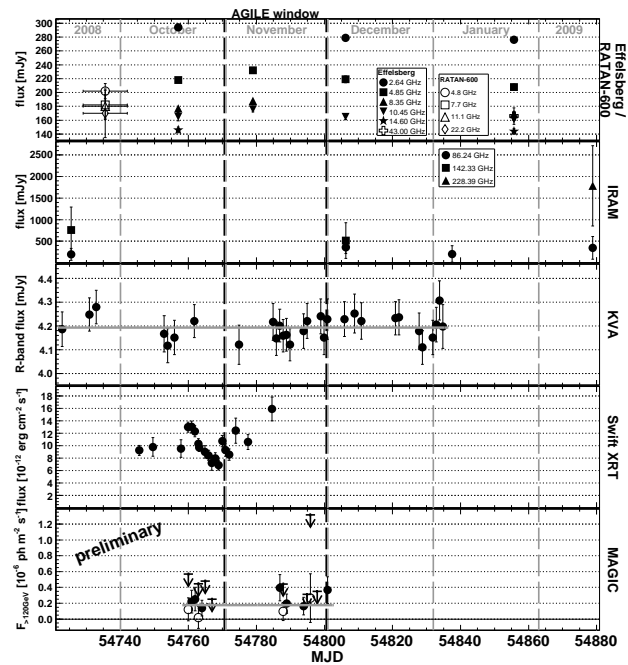}
\caption{MW light curve of 1ES\,2344+514 with daily binning but RATAN-600. The host galaxy contribution is not subtracted from the KVA data points. For MAGIC flux points consistent with or below 0, 99.7 \% c.l.\ upper limits are derived. For upper limits produced from a positive flux, the flux value is shown by an open circle. The grey solid lines represent linear fits to the optical and VHE data.}
\label{fig:LC_2344}
\end{figure}

The light curve of 1ES\,2344+514 (see Figure \ref{fig:LC_2344}) showed significant variability at radio (but IRAM) and X-ray wavelengths, whereas in the optical and VHE, variability could not be established (constant flux fit: $\chi^2 / d.o.f.$ = 13.1/29 and 5.5/14, respectively). Significant variability was also not apparent in the $Fermi$-LAT 1-year Catalog public light curve. The $Swift$ XRT measurements showed a steep flux increase by $\sim$ 50 \% within 2 days, followed by a rather slow decline, halving the flux in about 8 days. Another flare seemed to be present on the last day of observations, but missing coverage around that date prevented to draw any conclusions about the variability time scale. The radio fluxes were increasing towards MJD 54779.0, followed by a slow decrease, but were not obviously correlated with the optical fluxes. Correlated behaviour with other wavelengths could not be investigated due to the different light curve sampling. The VLBA results (one simultaneous observation on MJD 54762.0 was included in the long-term MOJAVE monitoring) are not shown in the figure as they probe much smaller regions of the source than the other radio measurements.% Their discussion is beyond the scope of this proceedings.

\begin{figure}[!t]
\centering
\includegraphics[clip, width=3.2in]{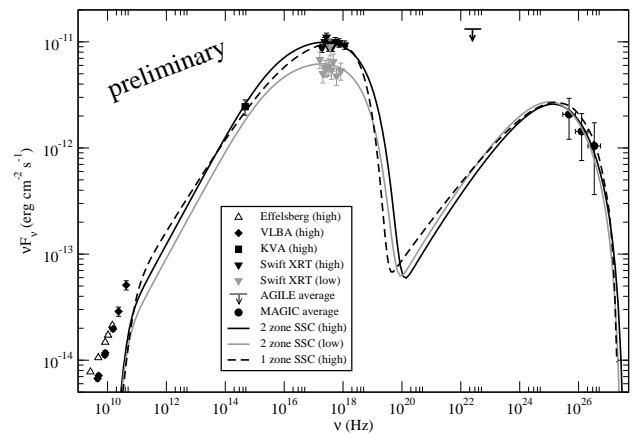}
\caption{SEDs of 1ES\,2344+514 resulting from this campaign. The KVA points are host-galaxy corrected, the MAGIC points are corrected for EBL absorption by \cite{Kneiske_lowerlimit}. The AGILE upper limit was calculated using a c.l.\ of 95 \%.}
\label{fig:SED_2344}
\end{figure}

Also in the case of 1ES\,2344+514, two nights with particularly low and high X-ray flux (MJDs 54766.9 and 54760.9, respectively) were chosen to construct simultaneous SEDs, as shown in Figure \ref{fig:SED_2344}. The MAGIC spectrum, averaged over all observation nights, was de-absorbed using the lower limit model of \cite{Kneiske_lowerlimit}. Note that due to lack of significant variability at VHE, the rather large error bars and only a factor of 2 difference in X-ray flux, these points give a good estimate both for the low and high state SED.

\section{Discussion and Conclusions}
Two different models have been applied to the simultaneous SEDs: a leptonic one-zone SSC model \cite{Tavecchio03} as well as a self-consistent two-zone SSC model \cite{Weidinger}. The physical conditions of the first one are defined by the parameters of the emitting source (magnetic field $B$, Doppler factor $\delta$ and source radius $R$) as well as the characteristics of the emitting electrons (with density $K$ at Lorentz factor $\gamma = 1$ and a broken power law distribution with spectral index $e_1$ between $\gamma_{min}$ and $\gamma_{break}$, and $e_2$ between $\gamma_{break}$ and $\gamma_{max}$). In the case of the second model, the parameters of the electron population are self-consistently derived from an initial injection of electrons with density $K$ into the acceleration region. Note that the model fits with \cite{Weidinger} shown in Figures \ref{fig:SED_Mrk180} and \ref{fig:SED_2344} are very preliminary and more refined modelling is ongoing. Due to the poor sampling of the 1ES\,2344+514 low flux state, only one model fit using \cite{Tavecchio03} has been produced. With more data arriving in the future, a dedicated fit will be presented also for the low flux state.

1ES2344+514 could be described well by both models in low as well as high flux state%, which is not surprising taking into account the rather low sampling of the SED. Further data will be added soon to better constrain the models
. The source was observed in one of the lowest flux states ever in VHE, X-rays and optical, whereas the radio fluxes remained on normal levels. Despite the low flux level, the derived model parameters (see Table \ref{tab:parameters}) are typical for HBLs. This means we either have not yet measured the quiescent state of the source, or the quiescent state parameters do not differ considerably from the standard parameters published up to now. The upper limits on size and magnetic field strength in the dominating radio emitting region derived from VLBA observations \cite{VLBA2344} do not contradict the parameters of the blazar emission zone as given in Table \ref{tab:parameters}.
\begin{table}[t]
\begin{center}
\begin{small}
\begin{tabular}{l||cc|cc||c|cc}
& \multicolumn{4}{c||}{Mrk\,180} & \multicolumn{3}{c}{1ES\,2344+514}\\\hline
& \multicolumn{2}{c|}{low} & \multicolumn{2}{c||}{high} & low & \multicolumn{2}{|c}{high}\\
& \cite{Tavecchio03} & \cite{Weidinger} & \cite{Tavecchio03} & \cite{Weidinger} & \cite{Weidinger} & \cite{Tavecchio03} & \cite{Weidinger}\\\hline\hline
$B$ & 0.11 & 0.21 & 0.11 & 0.08 & 0.10 & 0.07 & 0.10\\\hline
$\delta$ & 20 & 35 & 25 & 59 & 18 & 20 & 23\\\hline
$R$ & 3.6 & 1.5 & 1.0 & 2.3 & 6.0 & 3.7 & 6.0\\\hline\hline
$K$ & 1.3 &  & 8.2 &  & & 20  & \\\hline
$e_1$ & 1.8 & 2.3 & 1.8 & 2.2 & 2.0 &2.1 & 2.0\\\hline
$e_2$ & 4.0 & 3.3  & 4.0 & 3.2 & 3.0 & 3.1 & 3.0\\\hline
$\gamma_{min}$ & 10 & 7 & 10 & 7 & 7 & 1 & 7\\\hline
$\gamma_{break}$ & 5.0 & 2.5 & 30 & 15 & 3.3 & 8.0 & 3.5\\\hline
$\gamma_{max}$& 2.0 & 1.0 & 3.0 & 10 & 1.2 & 0.9 & 1.2\\\hline
\end{tabular}
\caption{Model parameters resulting from \cite{Tavecchio03} and \cite{Weidinger}. $B$ is given in units of G, $R$ in units of 10$^{15}$ cm, $K$ in 10$^3$ cm$^{-3}$, $\gamma_{break}$ in 10$^4$ and $\gamma_{max}$ in 10$^6$. $K$ of \cite{Tavecchio03} and \cite{Weidinger} are not directly comparable, hence not to be misleading it is given only for \cite{Tavecchio03}. See text for details.}\label{tab:parameters}
\end{small}
\end{center}
\end{table}

Also the SED corresponding to the low X-ray flux state of Mrk\,180 could be modelled well with standard parameters (see Table \ref{tab:parameters}). In contrast, the steep X-ray spectrum and rather high optical flux of the high state SED challenged the applied emission models. Both of them could not describe the data points sufficiently. \cite{Tavecchio03} greatly underestimated the optical flux, whereas \cite{Weidinger} required a high Doppler factor and assumed a flatter X-ray spectral index. One possible solution would be to consider separate emission zones for the X-ray and optical components. Note that during the high flux state, the synchrotron peak of the source shifted considerably from $\lesssim$ 0.5 keV to $\gtrsim$ 5 keV, which makes Mrk\,180 another extreme blazar candidate.

In conclusion, we organised the first MW campaigns from radio to VHE gamma-rays on the HBLs Mrk\,180 and 1ES\,2344+514. An unprecedented X-ray flare with a flux increase of $\sim$ factor 9 was detected for Mrk\,180, accompanied by a significant shift of the synchrotron peak to energies $>$ 5 keV. The applied SSC models could not fit the corresponding simultaneous SED satisfactorily, indicating that more emission regions or different physical mechanisms (like EC or hadronic models) are needed to reproduce the behaviour of the source. Variability could be established at all wavelength regimes with a hint of inter-band correlations. 1ES\,2344+514 was detected in an extremely low flux state in optical, X-rays and VHE gamma rays. The MW SED could be described well by SSC processes, yielding standard parameters for HBLs. Variability was present in radio and X-rays, whereas the flux at optical, high and very high gamma-ray energies was consistent with being constant. For both sources, data analysis and modelling is still ongoing.

\renewcommand{\baselinestretch}{0.7}\normalsize
\vspace{0.2cm}
\begin{scriptsize}
\textit{Acknowledgments.}
The MAGIC collaboration would like to thank the Instituto de Astrof\'{\i}sica de Canarias for the excellent working conditions at the Observatorio del Roque de los Muchachos in La Palma. The support of the German BMBF and MPG, the Italian INFN, the Swiss National Fund SNF, and the Spanish MICINN is gratefully acknowledged. This work was also supported by the Marie Curie program, by the CPAN CSD2007-00042 and MultiDark CSD2009-00064 projects of the Spanish Consolider-Ingenio 2010 programme, by grant DO02-353 of the Bulgarian NSF, by grant 127740 of the Academy of Finland, by the YIP of the Helmholtz Gemeinschaft, by the DFG Cluster of Excellence ``Origin and Structure of the Universe'', by the DFG Collaborative Research Centers SFB823/C4 and SFB876/C3, and by the Polish MNiSzW grant 745/N-HESS-MAGIC/2010/0.
The $Fermi$ LAT Collaboration acknowledges support from a number of agencies and institutes for both development and the operation of the LAT as well as scientific data analysis. These include NASA and DOE in the United States, CEA/Irfu and IN2P3/CNRS in France, ASI and INFN in Italy, MEXT, KEK, and JAXA in Japan, and the K.~A.~Wallenberg Foundation, the Swedish Research Council and the National Space Board in Sweden. Additional support from INAF in Italy and CNES in France for science analysis during the operations phase is also gratefully acknowledged.
The AGILE Mission is funded by the Italian Space Agency (ASI) with scientific and programmatic participation by the Italian Institute of Astrophysics (INAF) and the Italian Institute of Nuclear Physics (INFN).\\
We gratefully acknowledge N.\ Gehrels for approving this set of ToOs and the entire $Swift$ team, the duty scientists and science planners for the dedicated support, making these observations possible.
\end{scriptsize}

%\vspace{\baselineskip}

\renewcommand{\baselinestretch}{1.00}

\clearpage

\end{document}